\documentclass[aps,prl,preprint,showpacs,groupedaddress]{revtex4}
\usepackage {amssymb}
\usepackage {amsmath}
\usepackage {graphicx}
\usepackage {longtable}

\begin{document}

\title{Einstein's coefficients and the wave-particle duality \\
 in the theory of thermal radiation}

\author{Fedor V.Prigara}
\affiliation{Institute of Microelectronics and Informatics,
Russian Academy of Sciences,\\ 21 Universitetskaya, Yaroslavl
150007, Russia} \email{fprigara@imras.yar.ru}

\date{\today}

\begin{abstract}
It is shown that the concept of elementary resonator in the theory
of thermal radiation implies the indivisible connection between
particles (photons) and electromagnetic waves. This wave-particle
duality covers both the Wien and Rayleigh-Jeans regions of
spectrum.
\end{abstract}


\maketitle

The induced origin of thermal radio emission follows from the
relations between Einstein's coefficients for a spontaneous and
induced emission of radiation [1] (and references therein). The
strong argument in a favor of an induced origin of thermal
black-body radiation is that the spectral energy density in the
whole range of spectrum is described by a single Planck's
function. So if thermal radio emission is stimulated, then thermal
radiation in other spectral regions also should have the induced
character.

According to this conception, thermal emission from non-uniform
gas is produced by an ensemble of individual emitters. Each of
these emitters is an elementary resonator the size of which has an
order of magnitude of mean free path $l$ of photons,
$l=1/n\sigma$, where $n$ is the number density of particles and
$\sigma $ is the absorption cross-section.

The emission of each elementary resonator is coherent, with the
wavelength $\lambda = al$, where $a$ is a dimensionless constant,
and thermal emission of gaseous layer is incoherent sum of
radiation produced by individual emitters.

An elementary resonator emits in the direction opposite to the
direction of the density gradient. The wall of the resonator
corresponding to the lower density is half-transparent due to the
decrease of absorption with the decreasing gas density.

An elementary resonator can be considered as a realization of the
wave-particle duality. The wavelength of radiation emitted by a
resonator is determined by its size. On the other hand, the size
of an elementary resonator is determined by the mean free path of
photons. Thus, electromagnetic waves and particles (photons) are
indivisibly tied each with other in the concept of elementary
resonator.

This conclusion is contrary to the Einstein's opinion that the
energy fluctuations of thermal black-body radiation can be
attributed to the particles (photons) in the Wien region of
spectrum, and to the waves in the Rayleigh-Jeans region [2]. The
last statement has been obtained by application of the relation
between the probability and entropy, the universal validity of
which is a subject of ongoing debate [3].

[1] F.V.Prigara, Astron. Nachr., \textbf{324}, No. S1, 425 (2003).

[2] M.J.Klein, \textit{Einstein and the wave-particle duality},
Natural Philosopher, No. 3 (1964).

[3] V.Erofeev, in 12th International Congress on Plasma Physics,
25-29 October 2004, Nice, France, E-print archives,
physics/0409141.

\end{document}